\documentclass[aip,jap,reprint,amsmath,amssymb,amsfonts,longbibliography,floatfix
]{revtex4-1}

\usepackage{graphicx}
\usepackage{dcolumn}
\usepackage{bm}

\usepackage[utf8]{inputenc}
\usepackage[T1]{fontenc}
\usepackage{mathptmx}

\usepackage{hyperref}
\graphicspath{{figs/}}

\begin{document}

\title{Effect of tunneling on the electrical conductivity of nanowire-based films: computer simulation within a core--shell model}

\author{Irina~V.~Vodolazskaya}
\email{vodolazskaya\_agu@mail.ru}

\author{Andrei~V.~Eserkepov}
\email{dantealigjery49@gmail.com}

\author{Renat~K.~Akhunzhanov}
\email{akhunzha@mail.ru}

\author{Yuri~Yu.~Tarasevich}
\email[Corresponding author: ]{tarasevich@asu.edu.ru}

\affiliation{Laboratory of Mathematical Modeling, Astrakhan State University, Astrakhan, Russia, 414056}

\date{\today}

\begin{abstract}
We have studied the electrical conductivity of two-dimensional nanowire networks. An analytical evaluation of the contribution of tunneling to their electrical conductivity suggests that it is proportional to the square of the wire concentration. Using computer simulation, three kinds of resistance were taken into account, viz., (i) the resistance of the wires, (ii) the wire---wire junction resistance,  and (iii) the tunnel resistance between wires. We found that the percolation threshold decreased due to tunneling. However, tunneling had negligible effect on the electrical conductance of dense nanowire networks.
\end{abstract}

\maketitle

\section{Introduction}\label{sec:intro}

Nanowire networks (NWNs) have excellent optical, electrical and mechanical performances, which makes them very attractive for use as transparent electrodes in modern photovoltaics, light-emitting devices, touch screens, and thin-film transparent heaters.\cite{Manning2019SR} These applications require high transmittance of transparent electrodes while their sheet resistance should be within different ranges.\cite{Bae2012PhSc,Manning2019SR} The required sheet resistance varies from 500~$\Omega/\Box$ in the case of touch screens to 1~$\Omega/\Box$ in the case of solar cells.\cite{Bae2012PhSc} To ensure quality wire-to-wire contacts, different methods of treatment are used, viz.,  annealing,\cite{Lee2008NL,Madaria2010NR} plasmonic welding,\cite{Garnett2012NM} laser nanowelding,\cite{Han2014AM}  mechanical pressing,\cite{Tokuno2011NR}  electroless-welding,\cite{Xiong2016AM} capillary force,\cite{Lee2013AFM} the welding of crossed
silver nanowires (AgNWs) by chemically growing silver ``solder'' at the junctions of the nanowires,\cite{Lu2015AFM} capillary-force-induced cold welding,\cite{Liu2017NL} room-temperature plasma treatment,\cite{Zhu2013NT} electroplating welding,\cite{Zhao2019SEMSC} and electrical activation under ambient conditions using current-induced local heating of the junction.\cite{Bellew2015ACSN} Depending on the initial resistance, by using these technologies,the sheet resistivity can be reduced by several orders of magnitude to tens of ohms.

Depending on the manufacturing technology, the diameter and length of AgNWs may be different. Ref.~\onlinecite{Lee2008NL} reported on AgNWs that were $l = 8.7 \pm 3.7\,\mu$m long with a diameter of $d = 103 \pm 17$~nm, i.e., the aspect ratio, $\varepsilon$, was approximately $84$. In Ref.~\onlinecite{Nguyen2019NS}, the AgNWs had an average diameter of $d = 79 \pm 10$~nm and an average length of $l = 7 \pm 3\,\mu$m, i.e., $\varepsilon \approx 88$. In Ref.~\onlinecite{Bellew2015ACSN}, $l = 7 \pm 2\,\mu$m, $d = 42 \pm 12$~nm, i.e., $\varepsilon \approx 167$. In Ref.~\onlinecite{He2018JAP}, the mean length and diameter were $l = 66\,\mu$m $d = 160$~nm, respectively, i.e., $\varepsilon \approx 410$. Ref.~\onlinecite{Lee2016RSCA} reported on well-defined AgNWs with a narrow diameter distribution, uniform, through the range of 16--22~nm, with a long dimension of up to $20\,\mu$m, i.e., $\varepsilon \approx 1000$. AgNWs of the same aspect ratio $\varepsilon \approx 1000$ and lengths up to $200\,\mu$m have also been reported on in Ref.~\onlinecite{Xu2018JCIS}.  Thus, the aspect ratios of the AgNWs that have been considered range from 100 to 1000 in order of magnitude.

Four-probe measurements on almost 40 individual nanowires, with diameters ranging from 50 to 90~nm, gave an average resistivity of $20.3 \pm 0.5$~n$\Omega\cdot$m.\cite{Bellew2015ACSN} Junction resistance measurements of individual silver nanowire junctions have been performed and the distribution of junction resistance values presented.\cite{Bellew2015ACSN} The distribution of the junction resistance, determined using three different methods, showed a strong peak at 11~$\Omega$, corresponding to the median value of the distribution, and the presence of a small percentage (6\%) of high-resistance
junctions.\cite{Bellew2015ACSN} Thus, the junction resistance and the resistance of an individual  wire are mostly of the same order. Helpfully, Ref.~\onlinecite{Bellew2015ACSN} also presents a synopsis of previously reported junction resistance values.

Different approaches have been applied to compute the electrical conductivity of NWNs, e.g., a percolation approach,\cite{Zezelj2012PRB} an effective medium theory,\cite{Callaghan2016PCCP} a geometrical consideration,\cite{Kumar2016JAP} and a kind of mean-field approximation (where  transport within the system was described by the interaction of individual decoupled wires with a linear average potential background).\cite{Forro2018ACSN} A systematic analysis of the electrical conductivity of networks of conducting rods has recently been presented.\cite{Kim2018JAP}

Recently, percolating NWNs of widthless, stick nanowires have been considered.\cite{Benda2019JAP} The effective resistance of such NWNs has been studied by taking into account the wire resistance, wire---wire contact resistance, and metallic electrode---wire contact resistance. An accurate closed-form approximation of the effective resistance based on the above physical parameters has been proposed.\cite{Benda2019JAP}

An approximate analytical model for random NWN electrical conductivity has been proposed.\cite{Jagota2019Arxiv} This approach is faster than existing computational methods. The approach is based on the assumption that
$
\mathrm{E}[\sigma(\mathbf{M})] \approx \sigma(\mathrm{E}[\mathbf{M}]),
$
where $\mathrm{E}[\cdot]$ means the expected value, $\sigma$ is the sheet conductivity of a random NWN, and $\mathbf{M}$ is the adjacency matrix of this NWN.

A computational method has been developed to investigate the electrical properties of a silver NWN.\cite{Han2018SciRep} This method is based on extraction of the electrically conductive backbones and accounting for the wire---wire junction resistance.
An investigation has also been performed of how conductivity exponents depend on the ratio of the stick---stick junction resistance to the stick resistance for two-dimensional stick percolation.\cite{Li2010PRE}

To characterize NWNs, the number density, $n$, i.e., the number of wires, $N$, per unit area, may be used.\cite{Mertens2012PRE}
To mimic the shape of elongated particles and, at the same time, simplify connectivity simulations, different simple geometrical figures are used, e.g., sticks,\cite{Li2009PRE} rectangles,\cite{Li2013PRE} ellipses,\cite{Li2016PhysA} superellipses,\cite{Lin201917PT} and stadia.\cite{Tarasevich2019arXiv} A stadium is a rectangle with semicircles at a pair of opposite sides [Fig.~\ref{fig:intersections}(\textit{a})]. Its aspect ratio is
\begin{equation}\label{eq:asperctratio}
  \varepsilon = 1 + \frac{l}{2r}.
\end{equation}
A discorectangle (stadium) is the two-dimensional analog of a spherocylinder (a stadium of revolution or capsule), i.e.,  three-dimensional geometric shape consisting of a cylinder with hemispherical ends. When $\varepsilon = 1000$, the percolation thresholds for ellipses and rectangles are  $n_c \approx 5.624756$\cite{Li2016PhysA} and $n_c \approx 5.609947$,\cite{Li2013PRE}  respectively, while this value for zero-width sticks is $n_c \approx 5.6372858$.\cite{Li2009PRE} In such a way, real-world AgNWs ($\varepsilon \sim 10^3$) can, in simulations, be treated as zero-width sticks.

Although the presence of a giant component ensures the electrical conductivity of NWNs above the percolation threshold, electrical conductivity may, in principle, also occur slightly below the percolation threshold due to tunneling.\cite{Balberg2009JPhD} Moreover, tunneling may contribute to the electrical conductivity of NWNs even above the percolation threshold. In particular, there is an opinion that the conduction mechanism changes from tunneling to free electron conduction at increased AgNW concentration since the conductivity of the AgNWNs is mainly influenced by their surface and grain boundary scattering.\cite{Sohn2019Mat} Naturally, there should be a range of  concentrations where both the mechanisms are of approximately the same importance.

The tunnel conductivity between the two wires is defined as
\begin{equation}\label{eq:tonnel0}
  g^{(t)}_{ij} = g_0 \exp \left( -\frac{2\delta_{ij}}{\xi } \right),
\end{equation}
where $\delta_{ij}$ is the shortest distance between the $i$-th and $j$-th wires [Fig.~\ref{fig:intersections}(\textit{b})], $g_0$ is the conductivity of the junction between the two wires, while the tunneling decay length, $\xi$,  depends on the electronic potential barrier between the two wires.\cite{Ambrosetti2010PRB,Nigro2013PRE}
In fact, \eqref{eq:tonnel0} is a simplified notation of the Simmons' formula.\cite{Simmons1963JAP,*Matthews2018JAP}

Although the contribution of tunneling to the electrical conductivity of three-dimensional NWNs has been analyzed,\cite{Ambrosetti2010PRB,Nigro2013PRE,Nigro2014PRB,Haghgoo2019CPA} to the best of our knowledge, its contribution into the electrical conductivity of two-dimensional NWNs has not yet been fully studied. The goal of the present work was to obtain the dependencies of the electrical conductivity of NWNs  on their junction resistances. Two kinds of junction are involved in the consideration, viz., contact junctions and tunnel junctions. The rest of the paper is constructed as follows. In Section~\ref{sec:methods}, the technical details of the simulations and calculations are described and some tests are presented. Section~\ref{sec:results} presents our main findings. In Section~\ref{sec:concl}, we summarize and discuss the main results.

\section{Methods}\label{sec:methods}
\begin{figure}[!htb]
  \centering
  \includegraphics[width=\columnwidth]{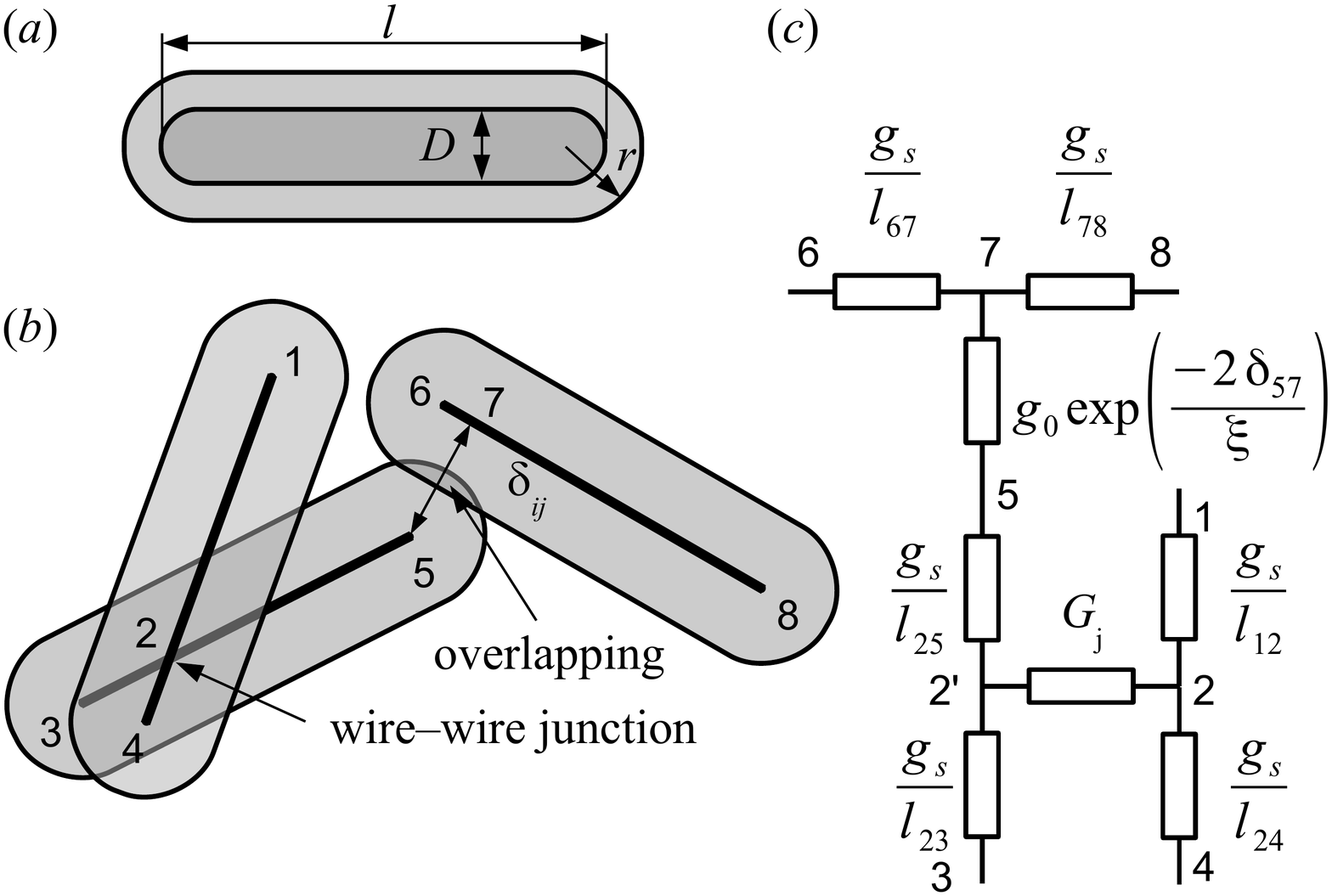}
  \caption{(\textit{a})~A core---shell model of a nanowire. (\textit{b})~Example of randomly deposited nanowires ($D=0$) (\textit{c})~Equivalent resistor network corresponding to the cluster of the three stadia shown in~(\textit{b}).\label{fig:intersections}}
\end{figure}

To simplify tunneling computations, the tunneling conductance is accounted only within the cutoff distance, $r$.
\begin{equation}\label{eq:tonnel}
  g^{(t)}_{ij} =
  \begin{cases}
    g_0 \exp \left( -\frac{2\delta_{ij}}{\xi } \right), & \mbox{if } \delta_{ij} \leqslant r,  \\
    0, & \mbox{otherwise},
  \end{cases}
\end{equation}
The region within the cutoff distance is a stadium.
In such a way, a wire consists of a core having the conductivity $g_s$, and a stadium shaped shell that can result in tunnel conductivity when two shells overlap [Fig.~\ref{fig:intersections}(\textit{a})].

Wires with $l =1$ and $D=0$ were added one by one randomly, uniformly, and isotropically onto a  substrate of size $L \times L$ having periodic boundary conditions (PBCs), i.e., onto a torus, until a  cluster spanning around the torus in two directions had arisen. Any two wires are supposed to be connected into one cluster when their shells (stadia) overlap each other [Fig.~\ref{fig:intersections}(\textit{b})]. $L=32$ was used in all our computations.  Since in our case the area is $L^2$, the desired number density is
\begin{equation}\label{eq:numdens}
  n = \frac{N}{L^2}.
\end{equation}
We denote as $n_c$ the number density corresponding to the occurrence of a spanning cluster.
We used the union---find algorithm\cite{Newman2000PRL,Newman2001PRE} to check for any occurrences of spanning clusters. In our study,  we used a version of the union---find algorithm adapted for continuous percolation.\cite{Li2009PRE,Mertens2012PRE} After the critical number density, $n_c$, had been determined, wires continued to be added until the desired concentration was reached.

The electrical conductivity of the spanning cluster was calculated as follows. For calculation of the electrical conductance, the PBCs were removed, i.e., the torus was unrolled into a plane, and two conducting buses were applied to the opposite borders of the system. The electrical conductance was calculated between these buses. Within the spanning cluster, all intersections of the cores and all overlappings of the shells were identified. The nearest distance between the cores of two stadia with overlapping shells may be either between their ends or between a core end of one of the stadia and an intermediate point on the core of the second one. We treated the latter as tunnel junctions. Both points of core intersections (contact junctions) and tunnel junctions divide the cores into segments. Each segment of a core between two junctions $i$ and $j$ of any of the two kinds was considered as a conductance $g_{ij} = g_s/l_{ij}$, where $l_{ij}$ is the length of the segment. Each core intersection was considered as providing an additional conductance $G_j$ while the conductances of tunnel junctions were calculated according to Eq.~\eqref{eq:tonnel}. In such a way, an electrical circuit was produced for each spanning cluster [Fig.~\ref{fig:intersections}(\textit{c})]. Having this electrical circuit, Kirchhoff's current law can be used for each junction of rods and Ohm's law for each occurrence of conductivity  between two junctions. The resulting set of equations was solved numerically to find the total conductance of the NWN. For each given  number of deposited stadia, $N$,  $100$ independent runs were performed to obtain the electrical conductance.

In all our computations, the electrical conductivity of the core was fixed, viz., $g_s = 1$~arb.u., while the tunnel electrical conductivity varied $ g_0 = 0.01 g_s, 0.1 g_s, g_s$.  We supposed that the case $G_j = g_0$ corresponds to untreated NWNs while $G_j = g_s$ corresponds to NWNs welded using the various technologies previously noted. Since the typical length of an AgNW is of the order $10^{-5}$~m while the tunneling decay length is of the order $10^{-9}-10^{-8}$~m,\cite{Nigro2013PRE,Ambrosetti2010PRB} we put $\xi/l = 10^{-3}$. Additionally, we studied the effect of the tunneling decay length on the electrical conductance for the ratios $\xi/l = 6\cdot 10^{-3}$ and $\xi/l = 10^{-2}$. The cutoff distance and the tunneling decay length were connected as $r = 1.25\xi$. The rationale for this choice is given in Section~\ref{subsec:evaluation}.

The error bars are of the order of the marker size when not shown explicitly. Due to isotropy, the electrical conductances in the two perpendicular directions are the same within statistical error, thus, only the conductivity in one direction is presented in all the figures.

Verification of the computer program was performed by using a comparison with known results for NWNs without tunneling.
The electrical conductance, $G$, of a dense random stick networks can be calculated as
\begin{equation}\label{eq:Zezelj}
  G = a \frac{(n-n_c)^t + c(L/l)^{-t/\nu}}{b n^{t-1}/G_s + (n + n_c)^{t-2}/G_j},
\end{equation}
where $a$, $b$  and $c$ are the fitting parameters, $t$ is the universal conductivity exponent, $n$ is the number density of fillers, $n_c$ is the percolation threshold, and $G_s$ and $G_j$ are the electrical conductivities of the sticks and junctions, respectively.\cite{Zezelj2012PRB} When~\eqref{eq:Zezelj} is written in terms of resistance instead of conductance, viz.,
\begin{equation}\label{eq:ZezeljR}
  R = \frac{b n^{t-1} R_s + (n + n_c)^{t-2}R_j}{a\left((n-n_c)^t + c(L/l)^{-t/\nu}\right)},
\end{equation}
the linear dependence of the NWN resistance on both the stick resistance, $R_s$, and the junction resistance, $R_j$, is clearly seen.

Based on the analysis of experimental data, a simpler relationship with only one fitting parameter, $\alpha$,
\begin{equation}\label{eq:Ponzoni}
  R \approx \alpha\left( R_n + R_j\right)
\end{equation}
has recently been proposed.\cite{Ponzoni2019APL} Here $R_n$ is the averaged value of the electrical resistance between two junctions.

Based on the geometrical consideration, another linear relationship has been proposed
\begin{multline}\label{eq:KumarR}
  R = \frac{\pi}{2\sqrt{N_E}} \left( \frac{4\rho}{\pi D^2} + \frac{R_j}{d}\right) = \frac{\pi}{2 d \sqrt{N_E}} \left( R_n + R_j \right)=\\ \frac{\pi}{2 d \sqrt{N_E}} \left( d R_s + R_j \right), 
\end{multline}
where $N_E = n(Cn + \exp(-Cn) - 1)$ is the total number of wire segments, $D$ is the wire diameter, $$
d= \frac{1 - \exp(-Cn)}{ Cn} - \exp(-Cn)
$$
is the mean segment length, and $C= 2/\pi$; the wire length, $l$, is assumed to be unit.\cite{Kumar2017JAP}

\begin{figure}[!hbt]
  \centering
  \includegraphics[width=\columnwidth]{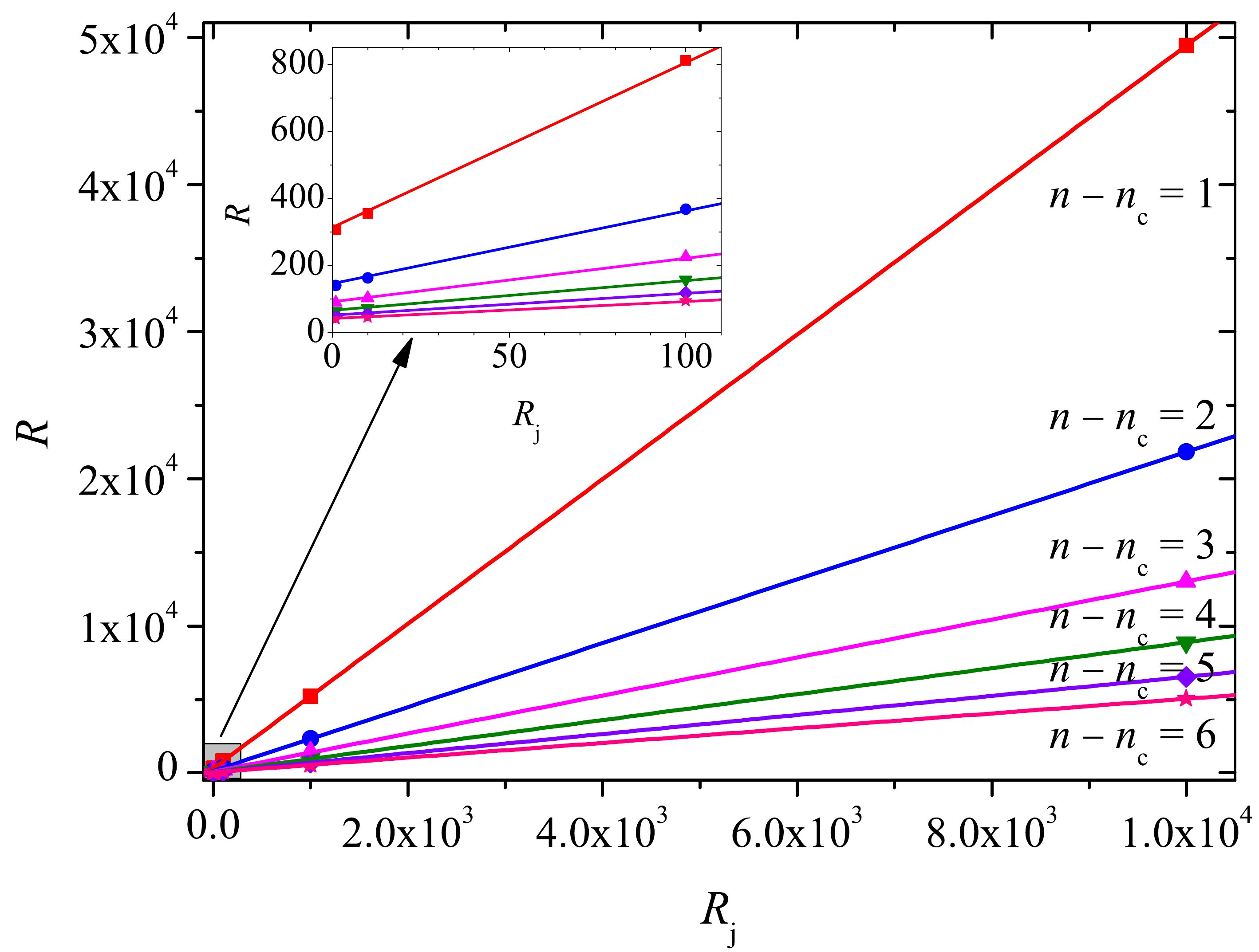}
  \caption{Dependency of the electrical resistance of NWNs, $R$, on the electrical resistance of their junctions, $R_j$, for different values of the number density, $n$, and fixed value of the cutoff distance $r = 0$. Here, $n_c = 5.79075$.}\label{fig:RvsRj}
\end{figure}

Recently, the linear dependence of the electrical conductance of NWNs on the wire resistance, junction resistance, and metallic electrode/nanowire contact resistance has also been proposed.\cite{Benda2019JAP} By contrast, a nonlinear dependency of the electrical conductivity of NWNs on both the wire resistance and the contact resistance has also been proposed.\cite{Forro2018ACSN} Figure~\ref{fig:RvsRj} demonstrates the linear dependence of the electrical resistance of the NWNs on the electrical resistance of junctions when tunneling is excluded from the consideration by setting the cutoff distance $r=0$.

\section{Results}\label{sec:results}
\subsection{Evaluation of the tunnel conductivity}\label{subsec:evaluation}
\begin{figure}[!htb]
  \centering
  \includegraphics[width=0.75\columnwidth]{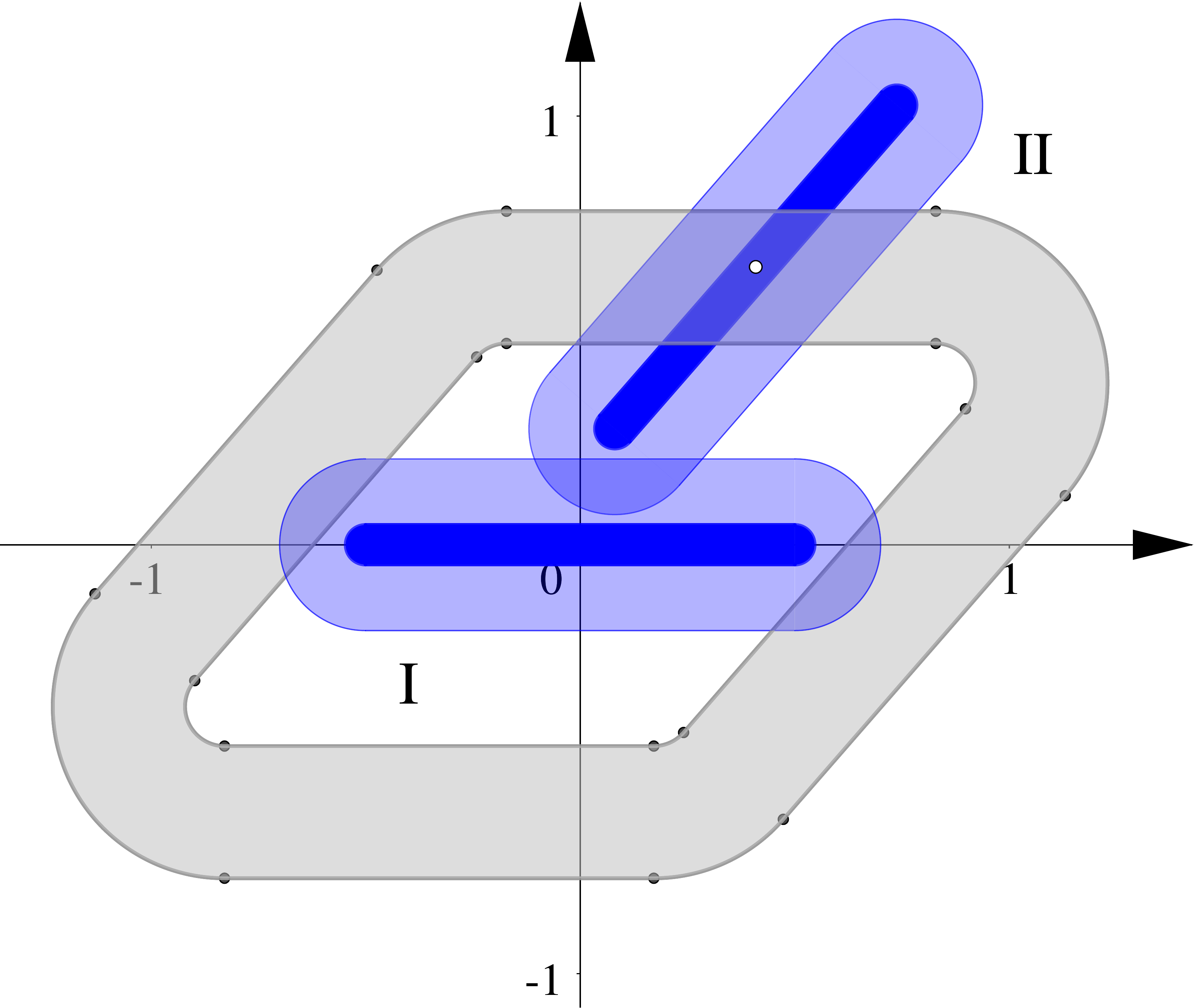}
  \caption{Wires I and II do not intersect while their shell overlap or touch each other when the center of wire II is situated within the shaded region.\label{fig:stadium}}
\end{figure}

We performed an evaluation of the contribution of tunneling to electrical conductivity.
By using a geometrical consideration (Fig.~\ref{fig:stadium}), one can conclude that two arbitrary wires I and II will not intersect each other  while their shells overlap or touch each other when the center of wire II is situated within the shaded region.
Note that the external perimeter of the shaded region limits the excluded area of the discorectangle.\cite{Balberg1984PRB}
The area of the shaded region is
$A = 4r(\pi r +  2 l ) - D(\pi D + 4 l).$ 
This conclusion is valid for arbitrary orientations of the wires.

The probability density function (PDF), that the shortest distance between two arbitrary wires equals exactly $\delta$ ($0 \leqslant \delta  \leqslant 2r$), is
\begin{equation}\label{eq:PDF}
 f(\delta) = \frac{2\pi (\delta + D)+ 4 l}{A}.
\end{equation}
The total contribution of all tunnel contacts to the electrical conductivity, $G_t = \sum_{i>j} g^{(t)}_{ij}$, is
$$
G_t = g_0 \sum_{\stackrel{i>j,}{\delta_{ij} \leqslant 2 r}} \exp\left(-\frac{2\delta_{ij}}{\xi} \right),
$$
The number of wires, for which their centers are located within the shaded area in Fig.~\ref{fig:stadium}, is $(N-1)A/L^2$. Hence, the number of pairs that obey the condition $\delta_{ij} \leqslant 2 r$, is
$$
\frac{N(N-1)A}{2L^2},
$$
the expectation of $G_t$ can be written as
$$
\langle G_t \rangle = g_0 \frac{N(N-1)A}{2L^2}\left\langle \exp\left[-\frac{2(\delta + D)}{\xi} \right] \right\rangle.
$$
The expectation can be evaluated as
 $$
 \left\langle \exp\left[-\frac{2(\delta + D)}{\xi} \right] \right\rangle = \int_{0}^{2r-D} f(\delta)  \exp\left[-\frac{2(\delta + D)}{\xi} \right] \, \mathrm{d} \delta.
$$
The integration yields
\begin{multline*}
\langle G_t \rangle =  \frac{ g_0\xi N ( N - 1 )}{4 L^2}\exp \left( -\frac{2D}{\xi}\right)\times \\
\left[ \pi\xi \left( 1 - \mathrm{e}^{-x} -x\mathrm{e}^{-x} \right) + (2 \pi D + 4 l)\left( 1 - \mathrm{e}^{-x}\right) \right],
\end{multline*}
where $x = (4r - 2D)/\xi$. Since $N \gg 1$,
\begin{multline}\label{eq:Gtcutoff}
\langle G_t \rangle =   g_0\xi n^2 L^2 l \exp \left( -\frac{2D}{\xi}\right)\times \\
\left[ \frac{\pi\xi}{4 l} \left( 1 - \mathrm{e}^{-x} -x\mathrm{e}^{-x} \right) +\left( 1 +\frac{\pi D}{2l}\right) \left(1 - \mathrm{e}^{-x}\right) \right].
\end{multline}
When $r \gg \xi$,
\begin{equation}\label{eq:Gt}
\langle G_t \rangle =   g_0 l \xi n^2 L^2 \exp \left( -\frac{2D}{\xi}\right)
\left( \frac{\pi\xi}{4l }+\frac{\pi D}{2l} + 1\right).
\end{equation}
When the cutoff distance is $1.25\xi$, the error in the value of $G_t$ obtained using Eq.~\eqref{eq:Gtcutoff} is about 1\%. This is the reason for our choice of $r = 1.25\xi$ for computations.

Since $\xi \ll l$ and $D \ll l$,
$\langle G_t \rangle  \approx  g_0 l \xi n^2 L^2 \exp \left( -\frac{2D}{\xi}\right)$.
In the particular case $D=0$, the electrical conductance is
\begin{equation}\label{eq:Gtapprox}
\langle G_t(n)\rangle =  g_0 N^2 \frac{\xi l}{ L^2}.
\end{equation}
Naturally, the contribution of tunneling into the electrical conductance is proportional to the square of the number of particles. In addition, the formula includes the ratio of the characteristic tunneling area $\xi l$ to the area of the entire system.

\subsection{Results of simulation}
\begin{figure}[!hbt]
  \centering
  \includegraphics[width=\columnwidth]{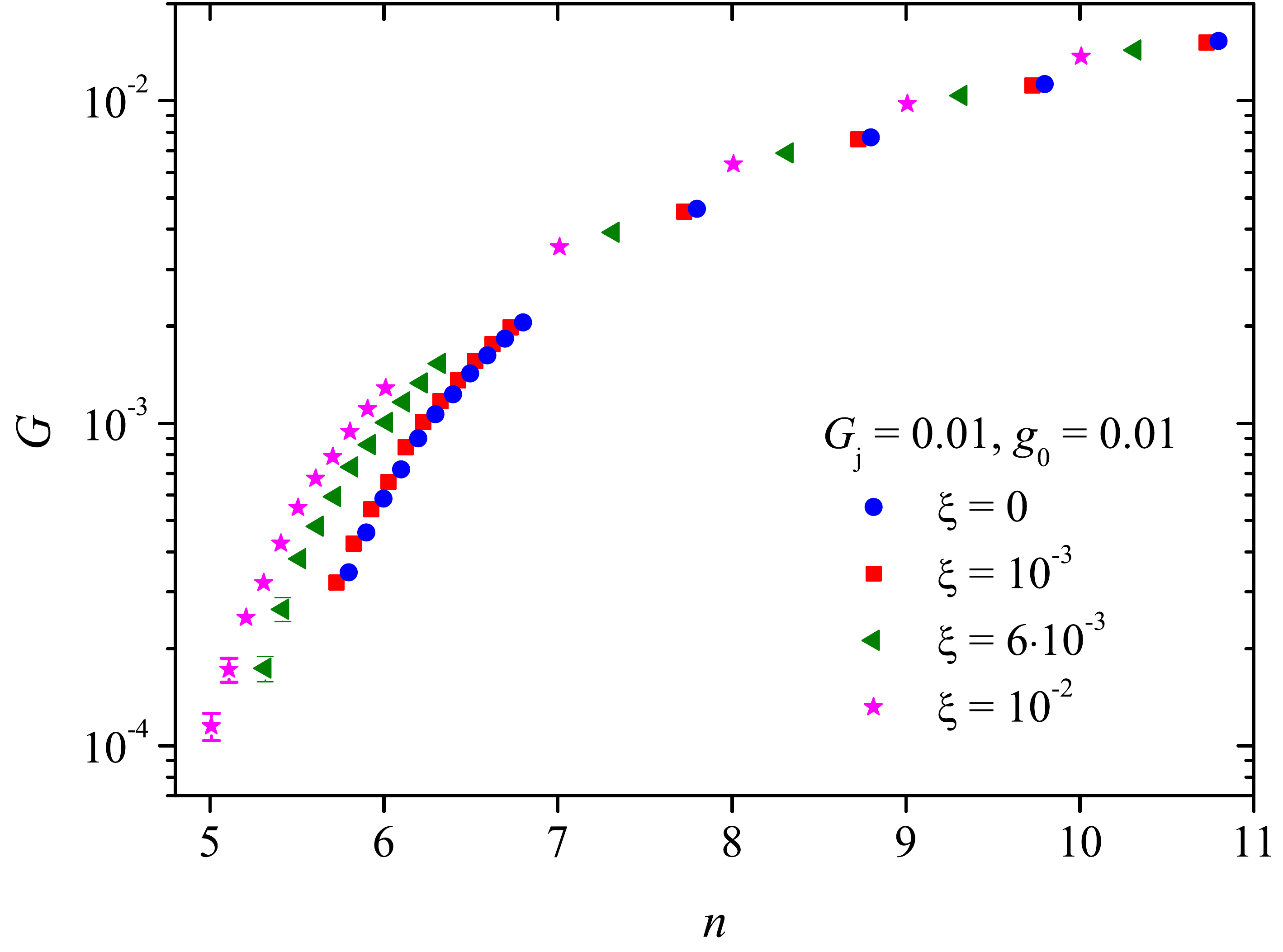}
  \caption{Dependency of the electrical conductance of NWNs, $G$, on the number density, $n$, for different values of the tunneling decay length, $\xi$, and particular values of the junction conductance, $G_j =0.001$, and $g_0 = 0.01$.}\label{fig:Gj-01g0-01}
\end{figure}

Figure~\ref{fig:Gj-01g0-01} demonstrates an example of the dependency of the electrical conductance of NWNs, $G$, on the number density, $n$, for different values of the tunneling decay length, $\xi$, and low values of the junction conductance, $G_j =0.01$, and $g_0 = 0.01$. This set of parameters can be considered as corresponding to untreated NWNs. For values of the number density $n \gtrapprox 8$, the effect of tunneling on the electrical conductance is negligible. However, the percolation threshold decreases due to tunneling.

\begin{figure}[!htb]
  \centering
  \includegraphics[width=\columnwidth]{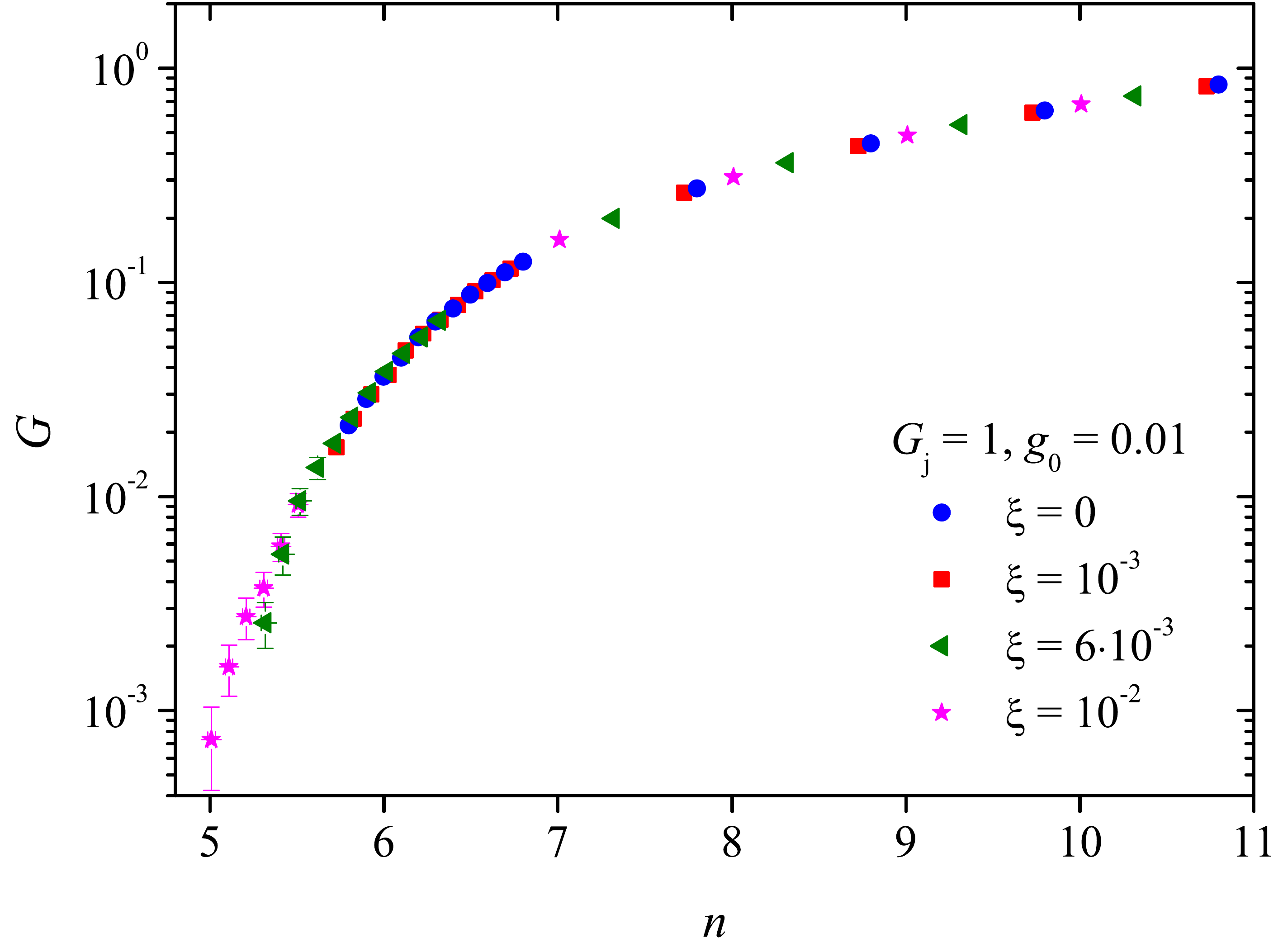}
  \caption{Dependency of the electrical conductance of NWNs, $G$, on the number density, $n$, for different values of the tunneling decay length, $\xi$. $G_j = 1$, $g_0 = 0.01$.\label{fig:Gj-1g0-01}}
\end{figure}

Figure~\ref{fig:Gj-1g0-01} shows the behavior of the electrical conductance of NWNs, $G(n)$,  for different values of the tunneling decay length, $\xi$, and high values of the junction conductance, $G_j = 1$ while the tunneling parameter is low $g_0 = 0.01$. This case can be treated as corresponding to welded NWNs since $G_j=g_s$. For values of the number density $n \gtrapprox 6$, the effect of the tunneling on the electrical conductance is negligible. However, the percolation threshold decreases due to tunneling.

\begin{figure}[!htb]
  \centering
  \includegraphics[width=\columnwidth]{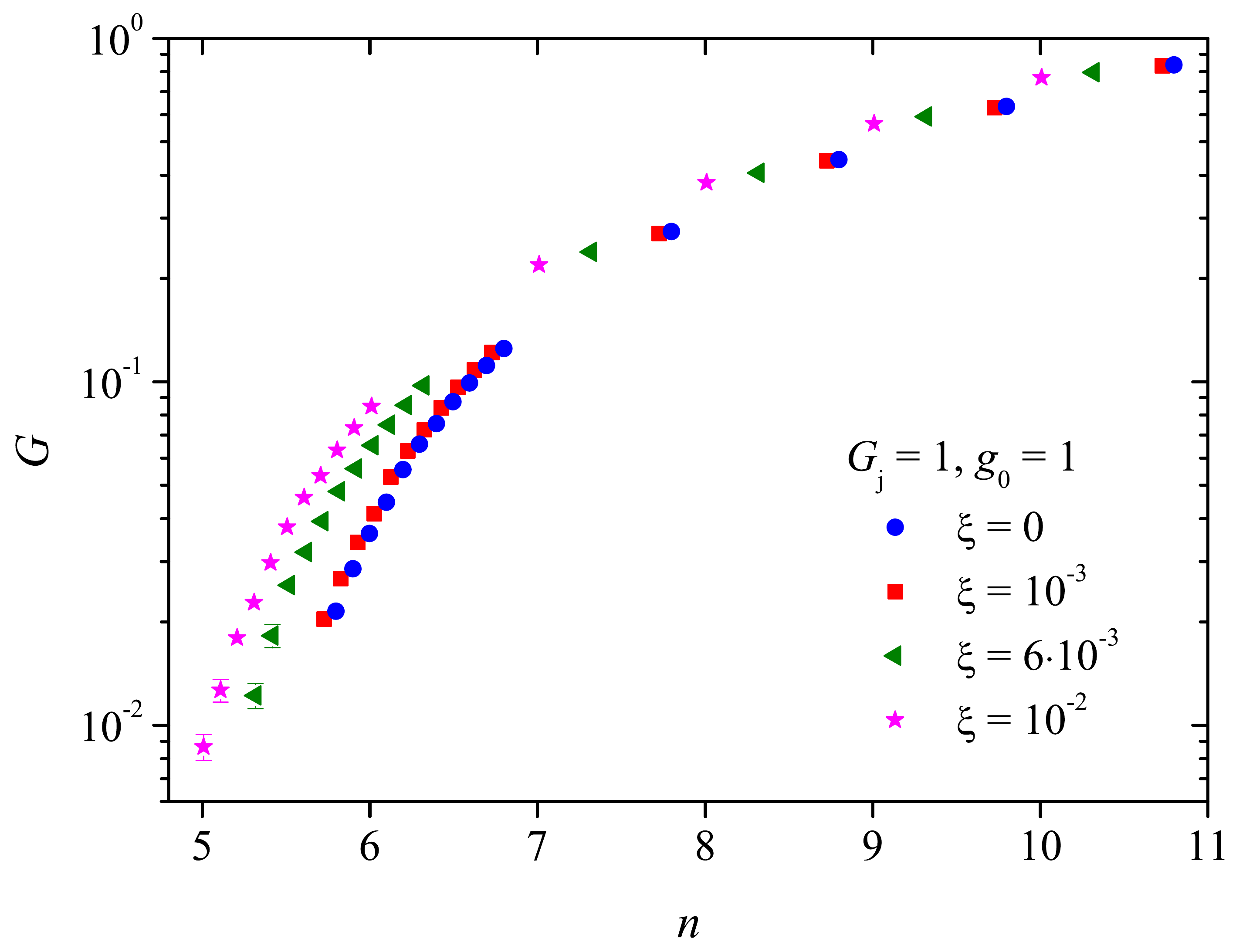}
  \caption{Dependency of the electrical conductance of NWNs, $G$, on the number density, $n$, for different values of the tunneling decay length, $\xi$, and high values both of the junction conductance and the tunneling parameter, viz., $G_j = 1$ and $g_0 = 1$.\label{fig:Gj-1g0-1}}
\end{figure}

Figure~\ref{fig:Gj-1g0-1} compares the dependencies of the electrical conductances of NWNs, $G(n)$, for different values of the tunneling decay length, $\xi$, and high values both of the junction conductance and the tunneling parameter, viz., $G_j = 1$ and $g_0 = 1$. This case can also be treated as corresponding to welded NWNs since $G_j=g_s$. The only difference from  Fig.~\ref{fig:Gj-01g0-01} is the larger magnitude of the electrical conductance.

\section{Conclusion}\label{sec:concl}

We have studied both analytically and numerically the impact of tunneling on the electrical conductivities of nanowire networks. A nanowire was represented as a hard conductive core with a soft shell that ensures tunneling occurs when shells belonging to different wires overlap. We found that the tunneling had a negligible effect on the electrical conductivity of dense networks. The  analytical consideration evidences that, in the slender rod limit (the aspect ratio of the wires tends to infinity), the contribution of the tunneling into the electrical conductance is  $g_0 N^2 \xi l/ L^2$.

\begin{acknowledgments}
We acknowledge the funding from the Ministry of Science and Higher Education of the Russian Federation, Project No.~3.959.2017/4.6.
\end{acknowledgments}

\bibliography{stadium}

\end{document}